\documentclass[epj]{svjour}
\usepackage{graphics}
\begin{document}
\title{Virtual Communities? The Middle East Revolutions at the Guardian Forum: Comment Is Free}
\author{Bernard Kujawski\inst{1} \and Peter Abell\inst{2}
}                     
\offprints{}          
\institute{Sandtable Ltd., London, UK \and MES Group, London School of Economics, London, UK}
\date{Received: date / Revised version: date}
%
\abstract{
We investigate the possibility of virtual community formation in an online social network under a rapid increase of activity of members and newcomers. The evolution is studied of the activity of online users at The Guardian -  \textit{Comment Is Free} forum - covering topics related to the Middle East turmoil during the period of $1^{st}$ of January 2010 to the $28^{th}$ of March 2011. Despite a threefold upsurge of forum users and the formation of a giant component, the main network characteristics, i.e. degree and weight distribution and clustering coefficient, remained almost unchanged. 
\PACS{
      {PACS-key}{discribing text of that key}   \and
      {PACS-key}{discribing text of that key}
     } 
} 

\authorrunning {B. Kujawski and P. Abell}
\titlerunning {The Middle East Revolutions at the Guardian Forum: Comment Is Free}

\maketitle
\section{Introduction}
\label{sec:intro}
The properties of social networks that are generated through the internet discussions of almost all forms, such as e-mails \cite{Barabasi-2005}, blogs \cite{Makowiec-2005}, discussion forums \cite{Kujawski-2007,palant}, bulletin board system \cite{Zhongbao-2003,Goh-2006} and more recently online social networks \cite{Golder-2007,Gilbert-2008} have been studied over last decade.

In recent years much interest has been expressed, particularly in the media, about the role which on line social networks may play in the mobilisation of social protest. The recent Middle East turmoil is no exception. However, there has been little direct documentation of the implied upsurge of network traffic occasioned by turbulent events.  This includes both those directly involved in such events and those who merely care to reflect upon the events.  

One of the most intriguing questions is the extent to which virtual communications might provide, at least in part, the foundations for a virtual community which, in turn, may eventually mobilise in the real world. It is beyond the ambitions of this paper to answer this very broad question but clearly it is pertinent to initially ask whether one can find evidence for the formation of virtual communities leaving it open as to whether the next step might or might not be taken.  This brief paper reports upon a study of the Guardian (a British daily newspaper) internet forum- Comment is Free- in order to ascertain the impact of the events precipitated by the social unrest and change of Government in Tunisia.

Standard network (graph and di-graph) theories \cite{Jackson-2009} would imply that tendencies towards community formation would be detected by the evolution of a high density of connections, transitive closure of triples, relative high node degrees and connection into a large component. 

\section{Data}
\label{sec:data}

A web crawler software was designed to download the complete content of both journalists' articles and public discussions occurring at the Guardian internet forum called \textit{Comment Is Free} (http://www.guardian.co.uk/commentisfree/all) during the period from the  $1^{st}$ of January 2010 to the $28^{th}$ of March 2011. 

Each day during that period articles appearing in The Guardian newspaper were published on-line and made available for public comment at the Comment is Free web-page. Every article was authored, dated and carried a keyword tagging its subject matter, e.g. \textit{UK Politics} or \textit{Barack Obama}.  

All those wishing to comment on any article are required to register; thus, all commentators are identified. Each comment carries its author's nickname and date of posting. In addition each comment can attract a number of recommendations by other commentators which provides an index of the popularity of any particular comment.  Comments always appear, in chronological order, directly under the main article and  can be registered for a period of up to four days after the article is posted, but not thereafter. 

 This paper focuses upon all the articles, which were tagged by one of  six keywords: \textit{Tunisia, Ben Ali, Egypt, Mubarak, Libya} and \textit{Gaddafi}. During the period studied, 279 journalists' articles were found that matched these tags, which then generated 35,942 comments posted by 6,217 identifiable commentators. 

\begin{figure}
\resizebox{0.45\textwidth}{!}{%
  \includegraphics{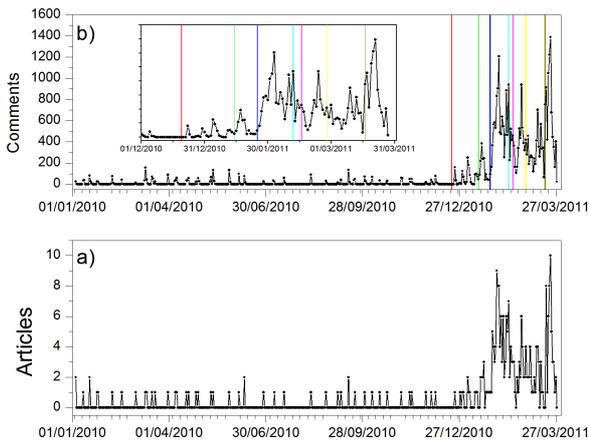}
}
\caption{The time series over the period of study of (a) the number of tagged published articles and (b) the number of comments. Inset in (b) is a zoomed series depicting greater detail. The vertical lines in (b) indicate, in chronological order, the major political events which are shown in Tab. \ref{tab1}}
\label{articles-comments}
\end{figure}

Fig. \ref{articles-comments}a depicts the time series of the number of appropriately tagged articles appearing in the forum and reveals a sharp upturn shortly after the beginning of 2011. 

\begin{table}
\caption{The major political events during the study period.}
\label{tab1}
\begin{tabular}{ll}
\hline\noalign{\smallskip}
Date & Event \\
\noalign{\smallskip}\hline\noalign{\smallskip}
20/12/2010 & Tunisia - Uprising of protest \\
14/01/2011 & Tunisia - Ben Ali fled the country \\
25/01/2011 & Egypt - Beginning of revolution \\
11/02/2011 & Egypt - Mubarak resigned from office \\
15/02/2011 & Libya - Gaddafi confronted peaceful protest \\
27/02/2011 & Egypt - PM Ghannouchi resigned \\
17/03/2011 & Libya - UN approved no-fly zone \\
\noalign{\smallskip}\hline
\end{tabular}
\end{table}

Fig. \ref{articles-comments}b, which depicts the frequency of posted comments over the specified study period, is accompanied by the date of certain major events (Tab. \ref{tab1}), post January of 2011. The insert accompanying Fig. \ref{articles-comments}b zooms in to the active period, giving a clearer picture of the incidence of comments. These time series reflect the upsurge in articles shown at Fig. \ref{articles-comments}a with a clear and precipitate increase in frequency of comments after about 1/1/2011, the occasion of the Tunisian uprising. The fluctuating increase in frequency seems also to be stimulated by each of the identified subsequent major events. The correlation of the time series in Figs. \ref{articles-comments}a and \ref{articles-comments}b, post 1/1/2011 is 0.88 (though, interestingly, it is 0.93 for the full period). This reflects the fact that before the upturn only a few immediate comments are provoked whereas afterwards the more frequent comments are spread over four days. Articles then, on average, cast a more sustained commentary shadow.

\section{Linking Commentators}
\label{sec:linking}

Internet discussion might appear to be an ideal candidate for a 2-mode network analysis \cite{Davis-1941,Newman-2001} whereby all the commentators who post comments on the same article are pair-wise connected. Such a network does not, though, in any way reflect interpersonal interactions, but only a common interest which might be interpreted as a necessary, though far from sufficient, condition for the existence of an embryonic virtual community. Such networks have, nevertheless, been used to infer interpersonal relationships \cite{Borgatti-1997}. However, the average number of distinct commentators per article in our dataset is 69 and the 2-mode network structures generated would have average degrees and network density parameters, way beyond those directly observed in interpersonal social networks. They are accordingly probably rather unreliable indicators of any on line social bonding.

In order to construct a directed (1-mode) social network structure, advantage can be taken of the fact that commentators typically attach to their comments the nickname associated with the comment towards which they wish to direct their remarks. The construct @nickname, has became an official standard among internet discussion users across many different internet discussion platforms.

	Thus, a weighted and directed link can be established between A and B when commentator A posts a comment that contains the nick name of commentator B which is attached to a prior comment concerning the same article. A weighting of the link is then computed as the number of times A comments on B.All links are directed and stamped by the time of the first direct communication by A directed to B. All commentators are also time-stamped, appearing in the evolving network for the first time when they comment upon another commentator.

\section{Results}
\label{sec:results}

\begin{figure}
\resizebox{0.45\textwidth}{!}{%
  \includegraphics{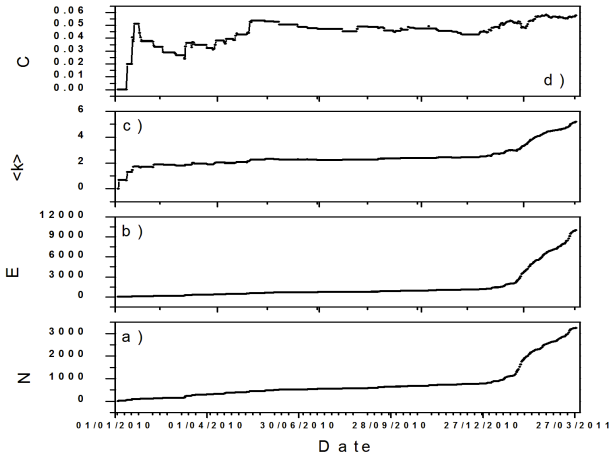}
}
\caption{Time series of (a) the number of nodes, (b) the number of directed edges, (c) the average degree $<k>$ (d) and the clustering coefficient.}
\label{nedc}
\end{figure}

The evolution of some major parameters of the evolving directed network, which my be interpreted as indicative of an emerging \textit{virtual community}, are depicted in Figs. \ref{nedc} a-d. An animation of the networks evolution can also be found at: http://www.youtube.com/watch?v=L7t-Fj-EAZs. 

Fig. \ref{nedc}a exhibits a sharp upturn in the number of commentators (i.e. the size of the node set) registering comments at the date already identified. Similarly, Figs. \ref{nedc}b and \ref{nedc}c show upturns in respectively the number of directed edges and the average degree of nodes (computed for the sake of simplicity as. the average number of links incident in and from a node) in the evolving structure. These are probably the results one would anticipate.

Fig. \ref{nedc}d depicts the movement of node clustering (i.e. C, the proportion of 3-cycles irrespective of the direction of links which are closed/transitive). Clustering is usually taken to be indicative of the formation of highly connected cliques or groups of nodes and, thus, community formation. Although there is a mild upturn in clustering at around 1/1/2011 there was also one a year before. There appears, therefore, to be only a slight tendency towards virtual clique formation at the start of 2011. The clustering coefficient hovers around 0.05 throughout the observation period and both maintains a low absolute value and fails to dramatically increase. Thus, despite the dramatic change in both the number of nodes (commentators) and edges after the start of 2011, there little support in these results for the view that a virtual grouping is in formation. This conclusion should, however, be treated with considerable caution.  There may be no need for a connected group of individuals to exhibit online clustering in order to provide a basis for a latent virtual community. It seems sensible to turn in the direction of connectivity.  
                  
In order to study the emergence of components all links were treated as undirected. That is to say any two nodes are deemed to be connected if there is a link in either or both directions.

\begin{figure}
\resizebox{0.45\textwidth}{!}{%
  \includegraphics{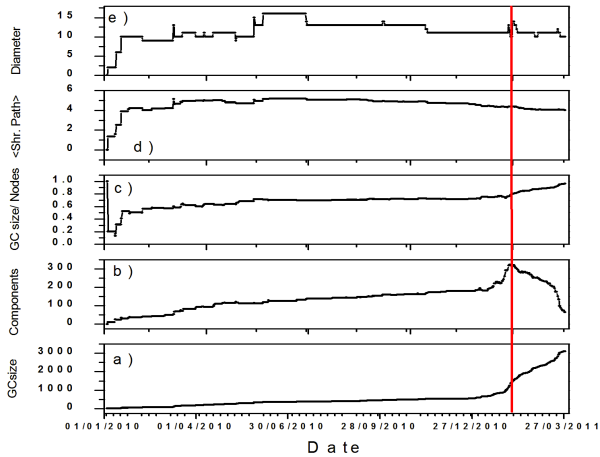}
}
\caption{Evolution of a Giant Component (GC), (a) its size, (b) overall number of components, (c) GC size / N ratio, (d) average shortest path in GC and (e) the diameter of GC.}
\label{giant-component}
\end{figure}

Fig. \ref{giant-component}a depicts the evolution of the size (i.e. number of nodes) of the giant (i.e. the largest) component; Fig. \ref{giant-component}b depicts the total number of components and Fig. \ref{giant-component}c the fraction of the nodes contained in the giant component. Figures \ref{giant-component}d and  \ref{giant-component}e show the time series of the average shortest path length and the network diameter, (both across components) respectively. The vertical line in all the diagrams indicates the date of maximum number of components (324) which was reached at 02/02/2011. After this date we observe a percolation connecting isolated components and at the termination of observation period 96\% of nodes belong to the giant component. These results strongly support the conclusion that quickened commentary leads to the emergence of a large component embracing most of the commentators. The average shortest path length declines very slightly but the emergence of a dominating large component appears to have little impact upon this parameter; likewise the diameter (i.e. the longest path in a component).   

\begin{figure}
\resizebox{0.45\textwidth}{!}{%
  \includegraphics{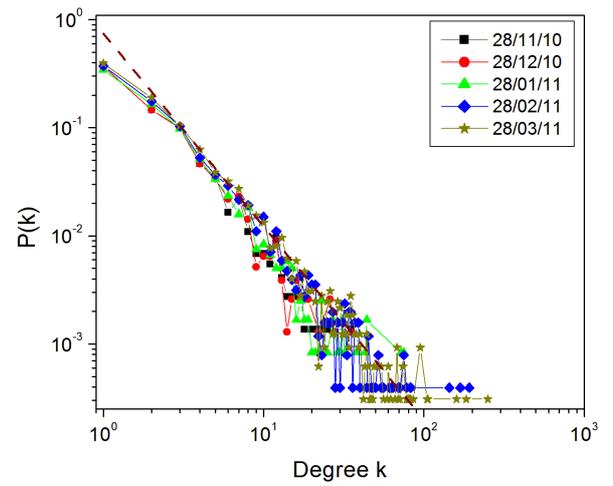}
}
\caption{Degree distribution in the revolution discussion network for the final 5 months. The distribution was fitted by power law distribution (red dashed curve) $f(x) = ax^{-\gamma}$, where $\gamma = 1.78 \pm 0.02$.}
\label{degree-distribution}
\end{figure}

Fig. \ref{degree-distribution} depicts the degree distributions across the existing nodes for different dates. The distributions appear remarkably consistent across time showing, at all the recorded dates, more variability at the higher levels of degree. Thus, although the number of directed links increases rapidly (Fig. \ref{nedc}b) after the start of 2011 the average degree does not correspondingly increase because of the increasing number of nodes (Fig. \ref{nedc}a) (i.e. commentators).      

\begin{figure}
\resizebox{0.45\textwidth}{!}{%
  \includegraphics{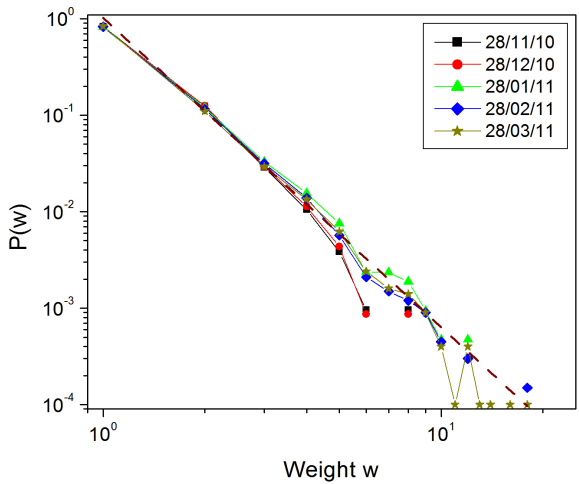}
}
\caption{The distribution of the directed edge weights for last 5 months of the observation period. The maximum link weight found is 18. The distribution was fitted by power law distribution (red dashed curve) $f(x) = ax^{-\beta}$, where $\beta = 3.20 \pm 0.03$.}
\label{weight-distribution}
\end{figure}

This conclusion is not altered by taking account of the weightings on the directed links (Fig. \ref{weight-distribution}). One might expect that, if a virtual community were to have been in formation, then the intensity of weightings would increase with the advent of the focal events. Again there is no support for the supposition.

The distribution of weightings at the end of the observation period is as shown at Tab. \ref{tab2}.

\begin{table}
\caption{The final percentage of edges with given edge weight.}
\label{tab2}
\begin{tabular}{cc}
\hline\noalign{\smallskip}
Weight	& \% of edges \\
\noalign{\smallskip}\hline\noalign{\smallskip}
1 & 83.0 \\
2 & 11.0 \\
3 & 3.0 \\
4 & 1.3 \\
5 & 0.6 \\
$>$ 5 & 1.1 \\
\noalign{\smallskip}\hline
\end{tabular}
\end{table}

These figures provide little support for the view that a virtual community is under formation. Very few commentators communicate with one another on more than two occasions. Moreover, the majority of edges (direct communication between two commentators) occurred only in respect of a single article; only 378 out of 10,021 edges (3.77\%) were found in more than one article. This again indicates little durability in the few relationships which are formed.

\section{Conclusions}

An analysis of the communication between commentators on a web site associated with a major national newspaper show an expected quickening of interest in articles relating to specified major international events. The question, thus, arises as to whether those involved might show signs of forming a virtual community. An inspection of clustering, degree distributions and the distribution of the weighting interpersonal communications offer little support for this proposition though a large component embracing most of the nodes is generated. Tab. \ref{tab3} gives a summary of the various properties of the evolved network at the termination of the study period. 

\begin{table}
\caption{Summary of the social network properties at the end of the observation period (28/03/2011).}
\label{tab3}
\begin{tabular}{lc}
\hline\noalign{\smallskip}
Property & Value \\
\noalign{\smallskip}\hline\noalign{\smallskip}
Nodes N	& 3239 \\
Edges E	& 10011 \\
Average degree $<k>$	& 5.18 \\
Clustering C	& 0.056 \\
Giant Component size	& 3109 \\
Components	& 64 \\
GC size / Nodes	& 0.96 \\
Modularity	& 0.334 \\
Communities	& 755 \\
Degree dist. Exponent $\gamma$	& 1.78 $\pm$ 0.02 \\
Weight dist. Exponent $\beta$ & 3.20 $\pm$  0.03 \\
\noalign{\smallskip}\hline
\end{tabular}
\end{table}

The lack of community formation may, however, be attributed to the nature of the discussion forum driven, as it is, by published journalists’ articles. On one hand commentators can easily comment upon an article, or direct their comment to another commentator, but on the other hand the web-page does not facilitate any other form of direct communication. A commentator has no other option than to browse through all articles and all associated comments to check if one of his/her \textit{friends} has posted a comment. Moreover, they cannot communicate directly and thus, depend entirely upon the frequency of postings by their favourite journalists. Closing the possibility of posting after 4 days can also reduce likelihood of furthering discussion and thus the generation of a virtual community. This lead to the conclusion that not every web page that provides users’ discussion place is capable of forming a virtual community. We identify two key elements that are missing at Comment Is Free forum and which seems to be vital from the perspective of an online forum user; a freedom of originating of a subject of a discussion and an effective way of managing online acquaintances. 

Whether the results documented here provide grounds upon which inferences about real communities can be made remains an open question. It will prove interesting to compare these results with a similar analysis appertaining to other notable events.  

\section*{Acknowledgements}

We are grateful to Andrew Skates for his help and support.

%

%
%

\end{document}